Artículo de investigación

# El Pensamiento Computacional en la Enseñanza de la Astrofísica Estelar: Análisis desde una intervención didáctica

*Computational Thinking in the Teaching of Stellar Astrophysics: Analysis from a Didactic Intervention*

Martínez Raba Duván Felipe [1]      duvan.martinez05@uptc.edu.co      https://orcid.org/0009-0007-5992-7635
Moreno Católico Angie Alejandra [2]  angie.moreno10@uptc.edu.co        https://orcid.org/0009-0002-3156-6894
Valderrama Daniel Alejandro [3]      daniel.valderrama@uptc.edu.co     https://orcid.org/0000-0002-3360-3890
Vargas Domínguez Santiago [4]        svargasd@unal.edu.co              https://orcid.org/0000-0002-5999-4842

Universidad Pedagógica y Tecnológica de Colombia (UPTC) | Tunja – Colombia | CP 510001

## RESUMEN

Este estudio diseña, implementa y evalúa una estrategia didáctica basada en el pensamiento computacional (PC) para fortalecer la enseñanza de la astrofísica estelar en educación secundaria. Se trabajó con 31 estudiantes de grado 11 de la Escuela Normal Superior Leonor Álvarez Pinzón (Tunja, Colombia) bajo un enfoque de investigación-acción. La metodología combinó análisis documental y aplicación práctica mediante una secuencia didáctica de cuatro talleres, integrando habilidades de PC como abstracción, descomposición y pensamiento algorítmico en actividades experimentales y modelado computacional. La evaluación, basada en rúbricas y matrices de desempeño, evidenció mejoras en la estructuración del pensamiento y la comprensión conceptual de fenómenos astrofísicos. Los resultados sugieren que el PC facilita la enseñanza de la física al promover un aprendizaje activo y estructurado. Se recomienda replicar la estrategia en distintos contextos educativos para validar su aplicabilidad y optimizar su impacto.

**PALABRAS CLAVE:** Astronomía, ciencias naturales, educación secundaria, habilidades de pensamiento computacional

## ABSTRACT

This study designs, implements, and evaluates a didactic strategy based on computational thinking (CT) to enhance the teaching of stellar astrophysics in secondary education. The research was conducted with 31 Year 11 students from Escuela Normal Superior Leonor Álvarez Pinzón (Tunja, Colombia) using an action-research approach. The methodology combined documentary analysis and practical application through a sequence of four workshops, integrating CT skills such as abstraction, decomposition, and algorithmic thinking into experimental activities and computational modelling. Assessment, based on rubrics and performance matrices, showed improvements in structured thinking and conceptual understanding of astrophysical phenomena. The findings suggest that CT facilitates physics education by promoting active and structured learning. It is recommended that this strategy be replicated in diverse educational contexts to validate its applicability and optimise its impact.

---

[1] Licenciado en Ciencia Naturales y Educación Ambiental, por la Universidad Pedagógica y Tecnológica Colombia - Colombia

[2] Licenciado en Ciencia Naturales y Educación Ambiental, por la Universidad Pedagógica y Tecnológica Colombia - Colombia

[3] Licenciado en Ciencia Naturales y Educación Ambiental, por la Universidad Pedagógica y Tecnológica Colombia - Colombia

[4] Doctor en Astrofísica, por el Instituto de Astrofísica de Canarias (IAC) - España

# INTRODUCCIÓN

La concepción y ejecución de la estrategia didáctica descrita en esta investigación fueron fundamentadas en los principios del Pensamiento Computacional (PC) como un enfoque didáctico innovador, orientado hacia el fortalecimiento de la comprensión de fenómenos astrofísicos en el contexto de las ciencias naturales. Este enfoque busca mejorar la enseñanza de la física mediante el desarrollo de habilidades cognitivas que faciliten la resolución de problemas complejos, la estructuración del pensamiento y la automatización de procesos (Wing, 2006). Sin embargo, en el ámbito de la educación secundaria, la integración del PC en la enseñanza de la astrofísica estelar sigue siendo incipiente, lo que limita su potencial como herramienta para mejorar el aprendizaje de conceptos fundamentales en esta área (Rojas y Vidal, 2022).

El problema identificado radica en la enseñanza tradicional de la física, que en muchos casos se basa en la transmisión de contenidos teóricos sin una conexión significativa con fenómenos observables, lo que genera dificultades en la apropiación de conceptos abstractos y en el desarrollo de habilidades analíticas en los estudiantes (González y Valderrama, 2021, Parra et al., 2021; Valderrama y González, 2024). En el caso de la astrofísica estelar, este desafío se amplifica debido a la complejidad inherente de los fenómenos involucrados, como la evolución estelar, la nucleosíntesis o la interpretación del espectro electromagnético de los astros (Rodríguez, 2018). Diversos estudios han señalado que la enseñanza de estos temas suele abordarse de manera descriptiva, sin aprovechar herramientas metodológicas que faciliten su comprensión desde un enfoque computacional (Valderrama, 2025; y Triana, 2024). Esto sugiere la necesidad de implementar estrategias didácticas que combinen la enseñanza de la física con el desarrollo de habilidades de PC, permitiendo una mayor interacción con modelos dinámicos y la solución de problemas mediante algoritmos y simulaciones.

En respuesta a estas limitaciones, la integración de contenidos de astrofísica estelar con estrategias basadas en el PC se plantea como una alternativa prometedora para fortalecer la enseñanza de la física en educación secundaria. El PC, entendido como un conjunto de habilidades cognitivas que permiten descomponer problemas complejos, formular soluciones algorítmicas y analizar datos de manera estructurada, ha demostrado ser una herramienta eficaz para la enseñanza de ciencias en distintos niveles educativos (Wing, 2006; Rojas y Vidal, 2022). En este sentido, su aplicación en el estudio de la astrofísica estelar facilita la interpretación de datos observacionales, el modelado de procesos físicos y la experimentación computacional con variables astronómicas (Cristóbal, 2017).

Para fundamentar la pertinencia de esta propuesta, se llevó a cabo una revisión sistemática en la base de datos académica Scopus, con el objetivo de identificar el estado del arte sobre el uso del PC en la enseñanza de la física y su aplicación en el aprendizaje de la astrofísica estelar. La búsqueda incluyó artículos publicados hasta el 30 de mayo de 2023 en inglés, español y portugués, utilizando términos como *"Computational Thinking"* AND *"Physics Education"*, *"Astrophysics"* AND *"Computational Thinking"*, y *"STEM Education"* AND *"Secondary School"*. Se establecieron criterios de inclusión que consideraron estudios que abordaran explícitamente la enseñanza de la física con estrategias de PC en educación secundaria, así como aquellos que analizaran su impacto en el desarrollo de habilidades analíticas y de resolución de problemas en contextos de ciencias naturales. Los criterios de exclusión descartaron investigaciones centradas únicamente en enseñanza de programación o en niveles universitarios sin aplicación en secundaria.

Los resultados de la revisión sistemática evidenciaron que la mayor parte de las investigaciones en PC se han centrado en áreas vinculadas a la informática y la tecnología, destacando su aplicación en programación y robótica (Martínez y Moreno, 2024). En contraste, se identificó una cantidad significativamente menor de estudios que establecen una conexión directa entre el PC y la enseñanza de la física en educación secundaria, lo que sugiere una brecha en la literatura existente. Además, la revisión mostró que, si bien existen estudios sobre la enseñanza de la astrofísica estelar (Vallejo, 2022), estos no han explorado en profundidad el uso del PC como estrategia didáctica para mejorar su aprendizaje. Esto subraya la necesidad de investigar cómo la integración del PC puede facilitar la apropiación de conceptos astrofísicos complejos y fortalecer el pensamiento crítico en los estudiantes.

En este contexto, la presente investigación se enfocó en diseñar e implementar una estrategia didáctica innovadora basada en PC para enriquecer la enseñanza de la física en educación secundaria, con énfasis en la astrofísica estelar. La estrategia enfatizó el desarrollo de habilidades específicas del PC, como el análisis de datos, la resolución de problemas, la depuración, la descomposición y el diseño de experimentos con enfoque algorítmico. Su aplicación en estudiantes de grado once de la Escuela Normal Superior Leonor Álvarez Pinzón, en Tunja, Boyacá, permitió evaluar el impacto del PC en la comprensión de los conceptos astrofísicos y su potencial para mejorar la enseñanza de la física en educación secundaria.

La enseñanza de la física juega un papel fundamental en la formación de ciudadanos críticos, capaces de comprender y analizar fenómenos científicos que impactan la sociedad (National Research Council, 2011). Ante los desafíos del siglo XXI, es imprescindible adaptar la enseñanza de esta disciplina a nuevos paradigmas que integren herramientas computacionales y enfoques interdisciplinarios. La astrofísica estelar, por su carácter altamente visual y conceptual, se presenta como un contexto ideal para explorar estas metodologías innovadoras, permitiendo a los estudiantes desarrollar una visión más estructurada de los procesos físicos y mejorar su capacidad de análisis mediante el uso del PC (Rodríguez, 2018; Vallejo, 2022). En este contexto, la presente investigación busca responder a la pregunta: ¿De qué manera las habilidades de pensamiento computacional favorecen el aprendizaje de la astrofísica estelar en estudiantes de grado once de la Escuela Normal Superior Leonor Álvarez Pinzón?, analizando el impacto de una estrategia didáctica basada en PC sobre la comprensión de conceptos astrofísicos y su potencial para transformar la enseñanza de la física en la educación secundaria.

## MATERIALES Y MÉTODOS

La investigación se desarrolló en la Escuela Normal Superior Leonor Álvarez Pinzón (ENSLAP), enfocándose en estudiantes de grado 11, el último año de educación secundaria. Se seleccionó una muestra de 31 alumnos, cuyas edades oscilaban entre los 16 y 18 años, representando un rango etario crucial previo al ingreso a la educación superior. La selección de participantes se realizó mediante un muestreo no probabilístico por conveniencia, dado que los contenidos curriculares de física en este nivel educativo se alineaban con los objetivos del proyecto. El grupo permitió analizar metodologías de enseñanza y evaluar prácticas didácticas para desarrollar el PC.

Desde el punto de vista investigativo, se adoptó el paradigma sociocrítico, que fomenta la reflexión crítica y busca transformar las dinámicas de enseñanza y aprendizaje mediante la educación (Unzueta, 2011). Este enfoque permitió reconfigurar la relación entre tecnología, sociedad y educación, promoviendo la participación de los estudiantes en la construcción de su propio conocimiento, especialmente a través de la integración del pensamiento computacional en el currículo de física.

La investigación empleó un enfoque cualitativo para interpretar y analizar los pensamientos y acciones de los estudiantes, estableciendo conexiones entre los aspectos sociales y los elementos investigativos (Cadena et al., 2017). Este enfoque permitió comprender el contexto natural en el que se desarrolló el estudio e interpretar los fenómenos educativos según los significados atribuidos por los participantes (Gurdián-Fernández, 2010). Así, se orientaron los esfuerzos hacia la transformación social mediante el análisis del potencial del PC para transformar y favorecer el aprendizaje de la astrofísica estelar, en línea con los objetivos planteados

El proyecto se enmarcó en la metodología de Investigación-Acción (I-A), la cual permite comprender y optimizar los procesos de enseñanza a través de reflexiones, planificaciones y prácticas innovadoras que mejoran progresivamente la educación (Bausela, 2004). Esta metodología ha demostrado ser eficaz en escenarios que promueven aprendizajes innovadores, facilitando la planificación y aplicación de proyectos educativos. Según Botella y Ramos (2019), la I-A resulta particularmente valiosa en modelos educativos no tradicionales, al permitir la enseñanza de conceptos complejos, especialmente en ciencias naturales como la física, fomentando así una transformación educativa significativa.

### Fases metodológicas

#### Fase de búsqueda y revisión documental

Se llevó a cabo una revisión sistemática en la base de datos Scopus utilizando las palabras clave "Computational AND Thinking". Los criterios de inclusión consideraron investigaciones que mencionaran el término "Pensamiento Computacional" en título, resumen o palabras clave; que estuvieran enfocadas en educación, específicamente en áreas como tecnología, informática, ciencias sociales o matemáticas; publicadas antes del 30 de mayo de 2023 y en idiomas español, inglés o portugués. Se excluyeron textos que no abordaron el pensamiento computacional en términos educativos, que no proviniera de contextos latinoamericanos o que no cumplieran con los criterios de inclusión establecidos.

**Fase diagnóstico y diseño**

Según la revisión sistemática, se identificó la necesidad de fortalecer los conceptos relacionados con las ciencias naturales, específicamente en física, principalmente en educación secundaria. Dado que esta área ha sido poco abordada desde el enfoque del pensamiento computacional. Para abordar esta problemática, se diseñó una secuencia didáctica compuesta por cuatro talleres.

La secuencia didáctica tuvo como objetivo fortalecer y facilitar la apropiación de conceptos físicos, mediante la astrofísica y las habilidades del PC. Para lograrlo, se estructuraron diversas actividades que despertaran el interés y motivación de las estudiantes, resaltando la importancia del conocimiento de aquellos fenómenos físicos en relación con la astronomía, desde su entorno y cotidiana.

Para cada taller se diseñaron herramientas didácticas (Figura 1), que permitieron interactuar con el estudiante, recolectando ideas sobre las temáticas abordadas, de acuerdo con la fase de la actividad. En la fase de diagnóstico Figura 1 a), se utilizaron preguntas al inicio de cada taller, con el fin de conocer conceptos y concepciones en física y astrofísica.

En cuanto a la fase de recopilación de datos e ideas, se emplearon diversas herramientas didácticas: "Export" Figura 1 b), diseñada para destacar aspectos curiosos de la astrofísica estelar y proporcionar un enfoque participativo para explorar datos interesantes relacionados con la astronomía; "Caja de Herramientas" Figura 1 c), cuyo objetivo fue proporcionar recursos que facilitaran el desarrollo de habilidades del PC, ofreciendo un conjunto de instrumentos conceptuales y prácticos; y "¡Hola Mundo!" Figura 1 d) como estrategia para que cada grupo sintetizara y expresara ideas concisas, conceptos o aportes fundamentales de las actividades realizadas en los talleres.

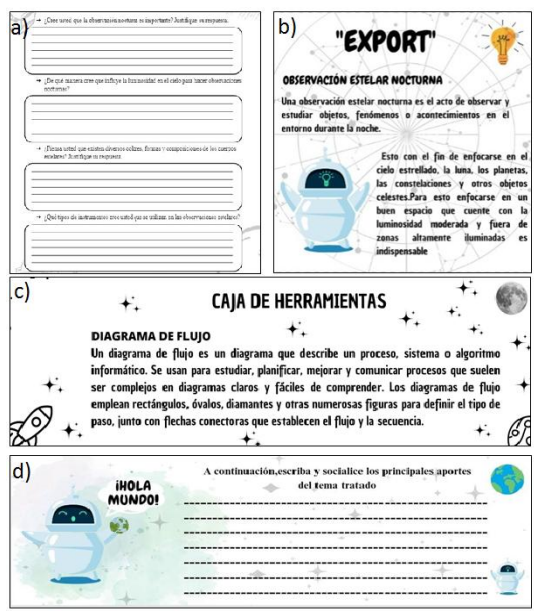

**Figura 1:** Herramientas didácticas

**Fase de implementación**

Según los resultados de las fases previas del estudio, se diseñó una secuencia didáctica cuyos contenidos propuestos se detallan en la Tabla 1. Esta secuencia didáctica se enfocó en fomentar las habilidades de PC, integrando para ello actividades que promueven el análisis de datos e información, el diseño de experimentos siguiendo enfoques algorítmicos, y la resolución de problemas a través de técnicas de descomposición y depuración. Además, se incentivó a los estudiantes a formular conclusiones e hipótesis empleando la generalización, lo cual permitió una comprensión más profunda de los conceptos tratados y el desarrollo de habilidades críticas de pensamiento, cada taller tuvo una duración promedio de dos sesiones cada una de dos horas de clase.

**Tabla 1.** Listado sobre talleres de la secuencia didáctica

| Taller | Conceptos y temáticas en Astrofísica Estelar | Conceptos Físicos | Habilidades de Pensamiento Computacional y Actividades Realizadas | Descripción de la Ejecución del Taller |
|---|---|---|---|---|
| Taller 1 | Astrofísica estelar observacional, Carta celeste, Diagrama H-R | Instrumentos ópticos, Nucleosíntesis, Fusión nuclear | Descomposición: División del análisis en dos fases: identificación de constelaciones con carta celeste y análisis evolutivo de estrellas con el Diagrama H-R. Pensamiento algorítmico: Construcción de diagramas de flujo para representar los procesos de observación y clasificación estelar. Abstracción: Identificación de patrones de evolución estelar en el Diagrama H-R y su relación con la composición de las estrellas. | Los estudiantes iniciaron con una fase de diagnóstico, respondiendo preguntas sobre observación nocturna. Posteriormente, trabajaron en la identificación de constelaciones con la carta celeste y exploraron su uso en la navegación. Luego, analizaron el Diagrama H-R, organizando estrellas según temperatura y luminosidad. Finalizaron diseñando diagramas de flujo para modelar el proceso de clasificación estelar y su evolución. |
| Taller 2 | Composición de estrellas | Longitud de onda, difracción, refracción, reflexión, espectro electromagnético | Automatización: Construcción de un espectrofotómetro casero para descomponer la luz. Depuración: Evaluación y corrección del diseño del espectrofotómetro para mejorar la precisión en la observación del espectro. Pensamiento algorítmico: Creación de infografías interactivas organizando los tipos espectrales estelares en patrones. Abstracción: Interpretación de los colores espectrales para inferir la composición química de las estrellas. | Se implementó una estrategia experimental, donde los estudiantes construyeron un espectrofotómetro casero utilizando CDs para analizar la descomposición de la luz en el espectro visible. Luego, identificaron las líneas espectrales en diferentes fuentes de luz. Se elaboraron infografías interactivas clasificando estrellas según su espectro, aplicando principios de reflexión y difracción. |
| Taller 3 | Termografía infrarroja | Espectro electromagnético no visible, longitud de onda y frecuencia, circuitos eléctricos, ley de Ohm | Descomposición: División de la actividad en tres partes: exploración térmica, modelado de circuitos y aplicación de la Ley de Ohm. Pensamiento algorítmico: Creación de reglas para interpretar imágenes térmicas y establecer relaciones con temperatura y emisión infrarroja. Abstracción: Relación entre los datos infrarrojos obtenidos y la teoría del espectro electromagnético no visible. | Se inició con una exploración térmica utilizando cámaras infrarrojas para detectar diferencias de temperatura en diversos materiales. Luego, los estudiantes diseñaron y probaron circuitos eléctricos, aplicando la Ley de Ohm para calcular resistencia y voltaje. Finalmente, se compararon imágenes térmicas y sus aplicaciones tecnológicas en astronomía y la vida cotidiana. |
| Taller 4 | Radioastronomía | Espectro electromagnético no visible, Magnetismo | Generalización: Extrapolación de la experiencia con detección de ondas de radio a la funcionalidad de los radiotelescopios. Pensamiento algorítmico: Diseño de procedimientos para captar y analizar señales de radio en el rango de 18 MHz a 22 MHz. Abstracción: Relación entre las ondas de radio detectadas y su origen astrofísico. | Se trabajó en la construcción de un detector de ondas de radio con materiales caseros, ajustando parámetros para captar señales en frecuencias entre 18 MHz y 22 MHz. Los estudiantes analizaron los datos recibidos, comparándolos con señales esperadas. Finalmente, discutieron la importancia de la radioastronomía y extrapolaron sus aplicaciones a la detección de cuerpos celestes. |

**Fase de evaluación**

Durante la secuencia didáctica, los estudiantes analizaron y discutieron conceptos de astrofísica estelar, documentando actividades con gráficos y textos. Esto permitió evaluar habilidades de PC y la comprensión de conceptos físicos en contexto. Además, este enfoque dual facilitó la valoración de los contenidos y la aplicación del razonamiento lógico y habilidades computacionales, integrando competencias científicas en el aprendizaje. Para evaluar las habilidades de pensamiento computacional, se utilizó un modelo de niveles que clasificó las competencias según su complejidad. Cada habilidad, como la abstracción, se dividió en tres niveles: "AB1" para un desarrollo mínimo, "AB2" para un nivel intermedio, y "AB3" para un nivel avanzado (Figura 2). Esta metodología, basada en Verdejo (2008), facilita identificar fortalezas y áreas de mejora en los estudiantes, proporcionando una visión detallada de su progreso en el desarrollo de estas competencias.

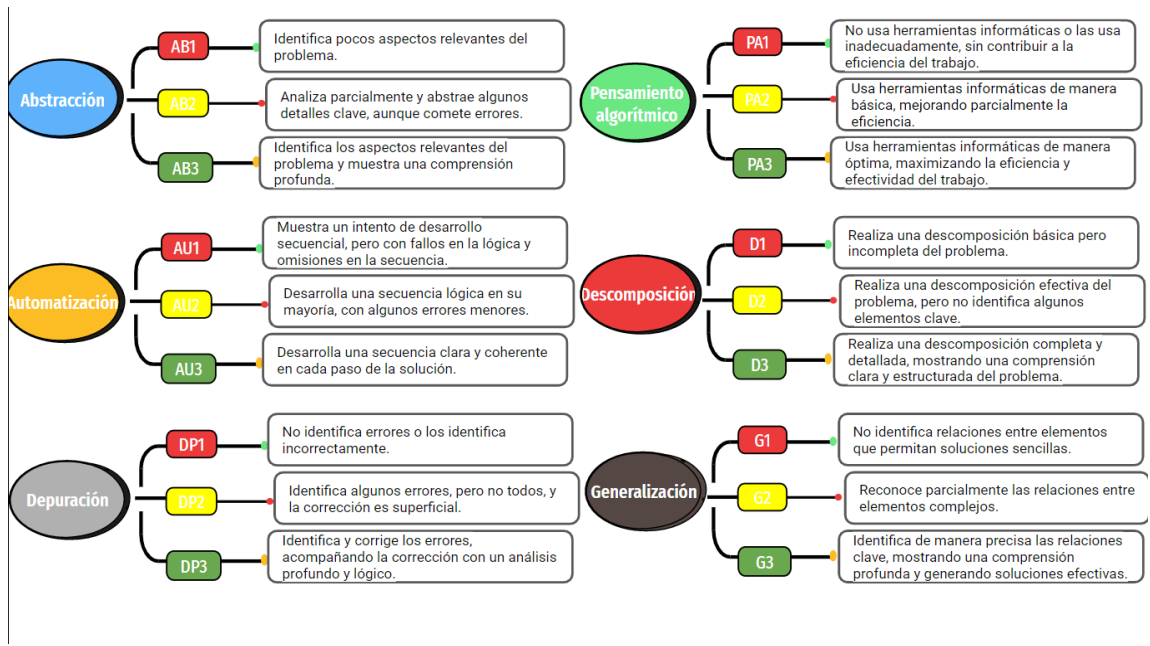

**Figura 2:** Matriz de evaluación de las habilidades del pensamiento computacional (PC).

La evaluación de los conceptos y temáticas en física y astrofísica se llevó a cabo mediante el diseño de una rúbrica de evaluación (Tabla 2). Esta elección se fundamenta en la aplicación de métodos de evaluación que se centran en criterios y estándares por niveles, los cuales determinan la ejecución de tareas, es decir el nivel de dominio de cada nivel y e objetivo alcanzado por el estudiante (Vera, 2008).

La rúbrica diseñada proporcionó un marco claro para evaluar y calificar el desempeño de los estudiantes en relación con los conceptos físicos y astrofísicos abordados en los talleres. Se establecieron criterios y niveles de logro de 0 a 3, siendo 0 el nivel más bajo y 3 el más alto, permitiendo una evaluación más objetiva y detallada de la comprensión de los estudiantes en estos tópicos.

**Tabla 2**: Rúbrica de evaluación conceptual y temática de astrofísica.

| TALLER | TEMAS | NIVEL 0 | NIVEL 1 | NIVEL 2 | NIVEL 3 |
|---|---|---|---|---|---|
| **Taller 1: Observación estelar nocturna** | Observación estelar nocturna | Identifica la observación estelar nocturna como una actividad recreativa, sin ningún tipo de beneficio a la sociedad. | Comprende el papel de la observación estelar nocturna como fuente de información de los cuerpos celestes. | Analiza la observación estelar nocturna como fuente de investigación y desarrollo social, cultural, tecnológico y científico. | Argumenta y describe el papel de la observación estelar nocturna, como centro que ha permitido el desarrollo de innovaciones sociales, culturales, tecnológicas y científicas. |
| | Carta celeste | No demuestra, ni identifica alguna comprensión básica de cómo leer una carta celeste. | Comprende el uso e idea de la carta celeste, pero se le dificulta el manejo de esta. | Aplica eficazmente el uso de la carta celeste para reconocer e identificar la posición de cuerpos celestes en fechas específicas. | Analiza patrones en los cuerpos celestes, al momento de comparar días o fechas en específico en la carta celeste. |
| | Diagrama H-R | No identifica, ni clasifica las diferencias entre las diferentes estrellas, de acuerdo con el color y tamaño. | Comprende las diferencias del tamaño y color de las estrellas, pero no logran relacionarlas con la materia y temperatura. | Explica el ciclo de vida de las estrellas, según el tamaño (material) y color (temperatura), de acuerdo con el ciclo evolutivo. | Analiza patrones en la distribución de colores y tamaños de las estrellas en las observaciones nocturnas. |
| **Taller 2: Espectro celeste visible** | Espectro electromagnético visible | No identifica colores del espectro electromagnético visible y algunos conceptos básicos. | Compara y explica la relación del espectro electromagnético con los diferentes cuerpos celestes. | Analiza el espectro electromagnético visible de los cuerpos celestes en función de la composición química de un cuerpo celeste. | Diseña y propone experimentos espectroscópicos para investigar cuerpos celestes específicos. |
| | Longitud de onda y frecuencia | No reconoce los conceptos de longitud de onda y frecuencia del espectro electromagnético | Comprende la diferencia de las distintas zonas del espectro electromagnético, de acuerdo con la longitud de onda y frecuencia, como también el color. | Hace uso de ejemplos que permitan diferenciar las diferentes zonas del espectro electromagnético relacionados con los cuerpos celestes. | Establece algún tipo de correlación entre la longitud de onda y la frecuencia, que permita identificar patrones en los cuerpos celestes. |
| | Percepción de la luz visible | No reconoce la conexión de manera básica, con limitaciones en la comprensión de cómo el ojo percibe los colores. | Comprende la función de la retina y cómo contribuyen a la percepción del color y la visión en condiciones de luz baja. | Explica la conexión entre el espectro visible y la percepción humana del color, destacando conceptos físicos y biológicos. | Diseña experimentos visuales, que permita apreciar los distintos colores que se observan en el espectro electromagnético |
| | Propiedades de la luz (reflexión, refracción, difracción) | No identifica las propiedades de la luz, ni los fenómenos que estas involucran. | Interpreta adecuadamente las principales propiedades de la luz, pero puede faltar profundidad o precisión en la explicación. | Emplea demostraciones y ejemplos acerca de las propiedades de la luz con la explicación teórica. | Diseña y construye experimentos, que permitan observar las principales propiedades de la luz. |
| **Taller 3: astrofísica de lo no visible** | Espectro no visible | No identifica ninguna región del espectro electromagnético no visible. | Comprende el espectro electromagnético no visible como sus distintas regiones y algunas aplicaciones. | Analiza las distintas regiones del espectro y comprende los parámetros longitud de onda y frecuencia, sin llegar a detallar en ellos, como también algunas aplicaciones. | Establece relaciones entre la longitud de onda y la frecuencia del espectro electromagnético no visible y señala aplicaciones en la vida cotidiana. |
| | Infrarrojo | No reconoce que la región del infrarrojo es un tipo de radiación electromagnética, originada por vibraciones de los átomos. | Identifica el infrarrojo como una región no visible en el espectro electromagnético, pero no tiene precisión en explicar su funcionalidad. | Analiza el infrarrojo como una región no visible en el espectro, como también algunas aplicaciones en la tecnología. | Establece relaciones entre la longitud y frecuencia del infrarrojo, resalta la importancia y algunas invenciones que se han desarrollado |
| | Circuitos eléctricos | No describe correctamente la funcionalidad de los circuitos eléctricos, ni describe sus funcionalidades. | Comprende adecuadamente los principios de circuitos eléctricos, incluyendo componentes básicos. | Ejecuta y construye de forma independiente circuitos eléctricos básicos; comete algunos errores menores. | Construye circuitos tanto mixtos, paralelos y en serie aplicando diseños, aplicando la ley de OHM. |
| **Taller 4: Radioastronomía** | Ondas de radio | No localiza en qué parte del espectro electromagnético se ubican las ondas microondas | Clasifica las ondas microondas como parte del espectro electromagnético sin reconocer su frecuencia y longitud de onda. | Identifica las ondas microondas en el espectro no visible gracias a su frecuencia y longitud de ondas, pero no define completamente su función. | Establece que la función de las ondas microondas es transmitir señales telegráficas gracias a su alto número de vibraciones por segundo, reconociendo su longitud de onda y frecuencia en las que se originan. |
| | Electromagnetismo | No reconoce lo que es el electromagnetismo como una rama de la física y su relación como rama de la física. | Ejemplifica el concepto de electromagnetismo con fenómenos magnéticos. | Identifica el electromagnetismo como una rama de la física que estudia fenómenos magnéticos y eléctricos. | Integra esta rama de la física para el análisis de los fenómenos eléctricos y magnéticos, su relación e interacción de las partículas de los campos magnéticos y eléctricos. |

# RESULTADOS Y DISCUSIÓN

**Revisión Bibliográfica**

Se analizaron 268 de 540 trabajos sobre pensamiento computacional disponibles en la base de datos Scopus, aplicando los criterios de inclusión y exclusión definidos. El análisis mostró que la mayor parte de las publicaciones se centraron en el nivel de pregrado, seguidas por primaria y, en último lugar, secundaria. Esta distribución evidencia una limitada intervención en la educación secundaria, reflejando un bajo nivel de apropiación del pensamiento computacional y un desarrollo mínimo de habilidades relacionadas en este nivel educativo (Figura 3).

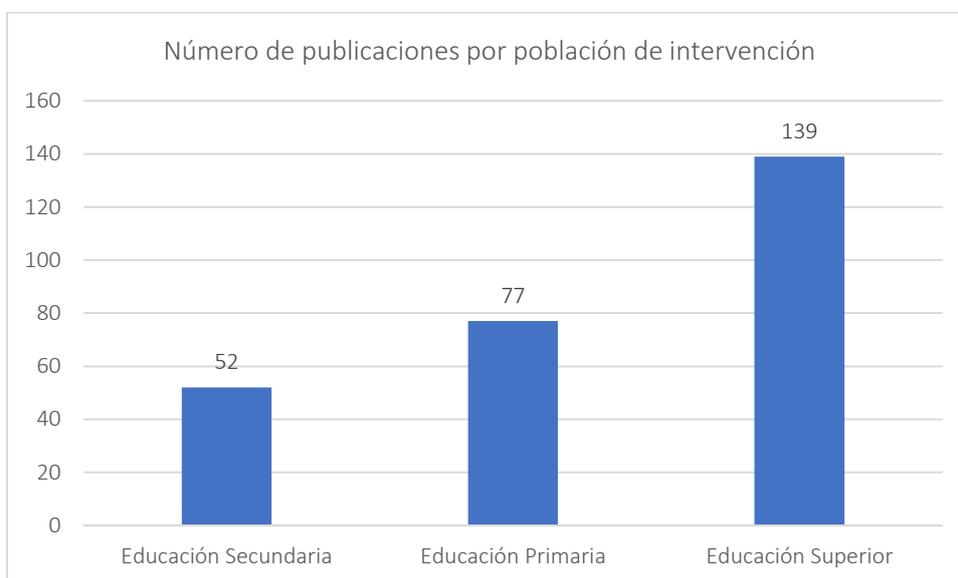

**Figura 3:** Distribución de las publicaciones por poblaciones de intervención.

En cuanto las relaciones disciplinares del PC (Figura 4), Tecnología e Informática lidera con 144 investigaciones centradas en programación y robótica, seguida por 82 trabajos sobre habilidades del PC. Matemáticas ocupa el tercer lugar con 27 estudios sobre cálculos vinculados al ámbito tecnológico. Ciencias Sociales aporta 10 investigaciones en temas geográficos y culturales, mientras que Ciencias Naturales registra 5 estudios enfocados en biología, química y física.

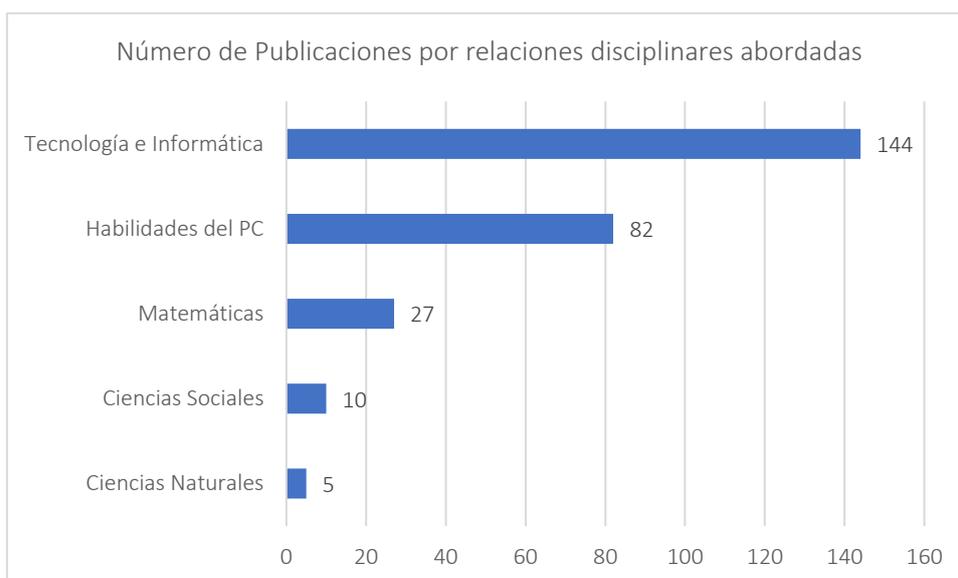

**Figura 4:** Distribución de las publicaciones por relaciones disciplinares abordadas desde el pensamiento computacional (PC)

Es importante destacar que la mayoría de los estudios se han centrado en la Tecnología e Informática, y en la propia integración del PC en el ámbito escolar. Por lo tanto, son pocas las relaciones interdisciplinarias y didácticas que se han planteado con otras disciplinas, siendo aún más escasas las investigaciones relacionadas con las Ciencias Naturales. En consecuencia, se recomienda a la comunidad científica ampliar las investigaciones en otras áreas del conocimiento, para tener un mayor campo de acción y romper el paradigma tecnológico del PC, explorando su potencial en diversas disciplinas y niveles educativos como en secundaria

## Intervención didáctica

A continuación, se presentan los resultados de la estrategia didáctica, que utiliza el PC como enfoque para mejorar las habilidades científicas en el aprendizaje de la física, específicamente en el campo de la astrofísica.

**Análisis de los resultados del Taller 1: Observación estelar nocturna**

En cuanto al Taller 1: "Observación estelar nocturna", el cual tenía como objetivo facilitar la observación estelar nocturna y desarrollar habilidades de PC en los estudiantes desde contextos astrofísicos. El taller constaba de cinco actividades, que incluían explorar conocimientos previos, diseñar diagramas de flujo, utilizar una carta celeste, identificar estrellas mediante el diagrama Hertzsprung-Russell (H-R) y clasificar estrellas según su color.

El análisis se enfocó en dos fases: las habilidades de PC y los aspectos conceptuales de la astrofísica estelar.

*Análisis de las habilidades del pensamiento computacional (PC)*

El análisis se centró en evaluar las habilidades del PC de abstracción (AB), pensamiento algorítmico (PA) y descomposición (D), a través de las actividades diseñadas en la estrategia didáctica como por los estudiantes en el Taller 1 (Figura 5).

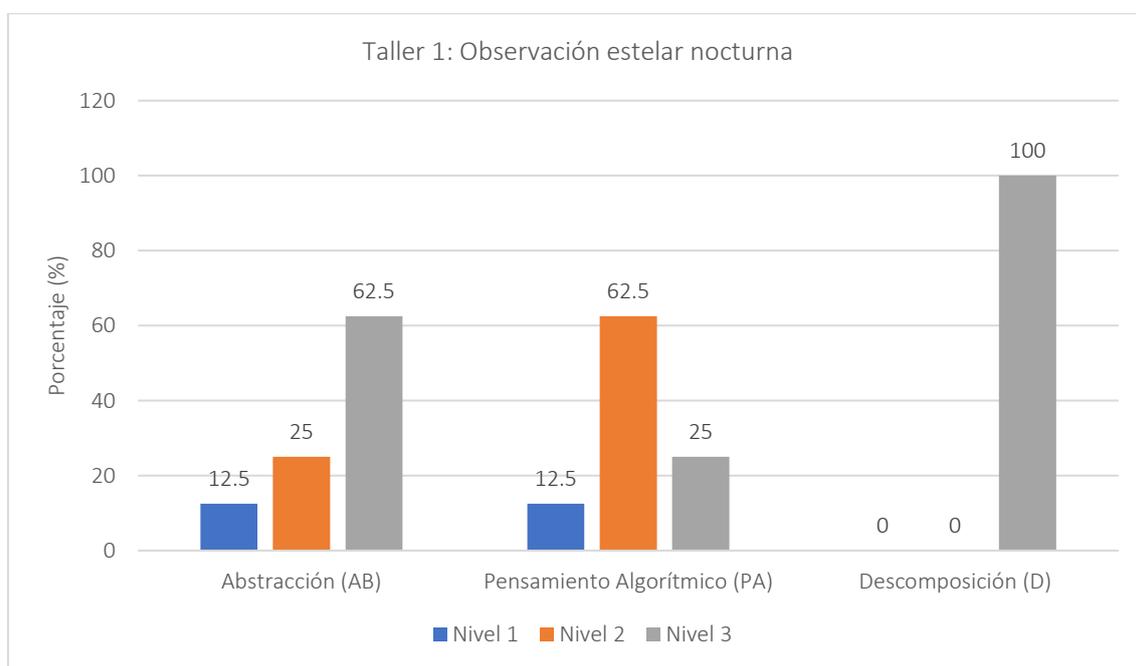

**Figura 5**: Niveles en las habilidades de pensamiento computacional (PC) del Taller 1

En cuanto a la habilidad de Abstracción, el 12.5 % de los grupos se ubicó en el nivel más bajo (AB1), identificando solo dos aspectos elementales de la observación estelar nocturna, lo que refleja una comprensión limitada de la actividad. En contraste, el 25% de los grupos alcanzó un nivel intermedio (AB2), reconociendo ciertos aspectos clave, pero sin articular una visión completa del proceso observacional, lo cual concuerda con Wing (2006), quien señala que la abstracción implica omitir detalles innecesarios sin perder la esencia del problema. Por otro lado, el 62.5% de los grupos logró el nivel más alto (AB3), al clasificar de manera estructurada los elementos fundamentales para realizar una observación estelar nocturna efectiva. Estos hallazgos concuerdan con Hernández et al. (2014), quienes resaltan que la abstracción es clave en la estructuración de problemas complejos al permitir la identificación de patrones esenciales sin perder de vista la globalidad del fenómeno estudiado.

En relación con el Pensamiento Algorítmico, el 25% de los grupos alcanzó el nivel más alto (PA3), evidenciando la capacidad de estructurar una secuencia lógica clara en sus diagramas de flujo. Este resultado es coherente con estudios previos que destacan que los organizadores gráficos, como los diagramas de flujo, facilitan la síntesis y representación estructurada de actividades, promoviendo la organización lógica de los procesos. Sin embargo, el 62.5% de los grupos se ubicó en el nivel intermedio (PA2), presentando errores secuenciales en sus diagramas, lo que sugiere que, aunque identificaron los pasos clave, no lograron estructurar de manera efectiva la resolución del problema, como lo señala Hernández et al. (2014), quien advierte que los errores en la secuencia pueden afectar la comprensión del proceso. Además, el 12.5% de los grupos permaneció en el nivel más bajo (PA1), evidenciando dificultades para desarrollar un hilo conductor coherente, lo que puede estar relacionado con la falta de experiencia en el uso de representaciones gráficas para modelar procesos

Con respecto a la habilidad de Descomposición, el 100% de los grupos alcanzó el nivel más alto (D3), demostrando la capacidad de dividir el problema en partes más pequeñas hasta llegar a la solución: clasificar una estrella según su color. Estos resultados refuerzan lo planteado por Wing (2006), quien enfatiza que la descomposición de problemas facilita su abordaje y resolución al segmentar un desafío complejo en componentes más manejables. Asimismo, la actividad permitió a los estudiantes identificar cómo la segmentación de características estelares (color, temperatura y tamaño) se correlaciona con los criterios científicos establecidos para la clasificación estelar, lo que se alinea con los principios de enseñanza de la astrofísica estelar y la metodología del diagrama de Hertzsprung-Russell (H-R)

.

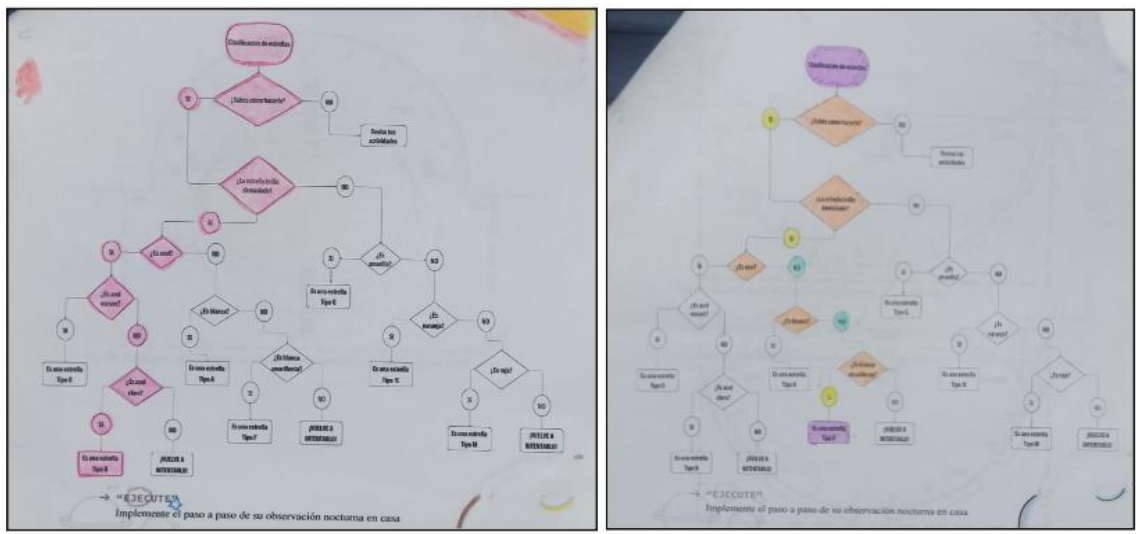

**Figura 6**: Diagramas de flujo. Clasificación de las estrellas Grupo 1 y Grupo 5.

*Fortalecimiento conceptual*

El análisis de los conceptos y temáticas de astrofísica estelar del Taller 1 se realizó a través de tres contenidos principales: la observación estelar nocturna, el uso de la carta celeste y el diagrama H-R (Figura 7).

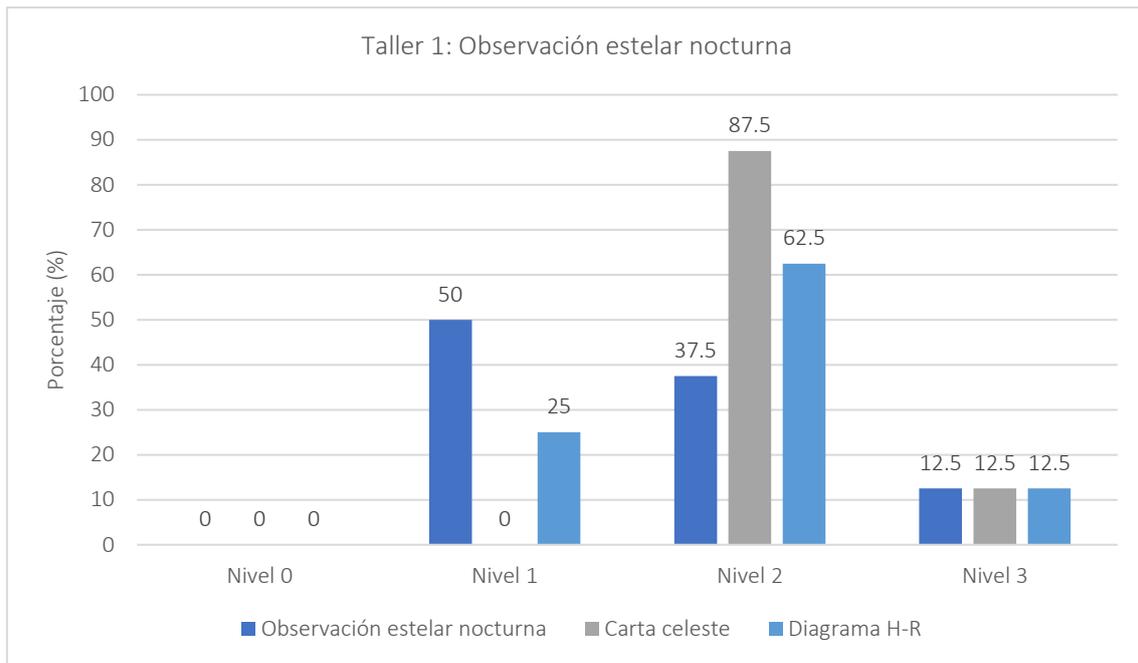

**Figura 7**: Niveles en la apropiación conceptual y temática del Taller 1

En el taller inicial, la observación estelar nocturna evidenció distintos niveles de apropiación conceptual. El 12.5% de los grupos alcanzó el nivel más alto (nivel 3), argumentando la relevancia de esta práctica en la astronomía, la navegación y su impacto en la sociedad. Este grupo estableció conexiones entre la observación estelar y su utilidad en la orientación marítima, en concordancia con Surroca (2019), quien resalta el uso de los astros como referencia para la navegación mediante instrumentos como el astrolabio. El 37.5% se situó en el nivel intermedio (nivel 2), identificando fenómenos astronómicos como cometas, asteroides y eclipses, pero sin integrar estos conocimientos en un contexto más amplio. El 50% restante permaneció en el nivel más bajo (nivel 1), limitándose a considerar la observación estelar nocturna como una actividad exclusivamente informativa sobre cuerpos celestes, sin reconocer su influencia en el desarrollo social, cultural y tecnológico.

El análisis del manejo de la carta celeste evidenció que el 12.5% de los grupos alcanzó el nivel conceptual 3, logrando comparar orientaciones y constelaciones en distintas ubicaciones geográficas. Este nivel de comprensión está alineado con Justiniano y Botelho (2016), quienes destacan la carta celeste como un recurso esencial en la identificación de objetos astronómicos y su evolución en el tiempo, desde su uso en la navegación hasta su aplicación didáctica en la enseñanza de la astronomía. En el nivel intermedio, el 87.5% de los grupos identificó constelaciones básicas y utilizó la carta celeste para reconocer cuerpos celestes en fechas específicas, sin analizar patrones comparativos más avanzados.

En el análisis del diagrama Hertzsprung-Russell (H-R), el 12.5% de los grupos estableció patrones entre el color, tamaño y propiedades de las estrellas, relacionando estas variables con su evolución estelar (nivel 3). Percy (2012) señala que la relación entre temperatura y color estelar permite comprender la secuencia evolutiva de las estrellas, lo cual coincide con los hallazgos de este grupo. El 62.5% de los grupos alcanzó el nivel 2, explicando el ciclo estelar a partir de tamaño y color, pero sin profundizar en variables adicionales como la composición química o la luminosidad. El 25% restante se mantuvo en el nivel 1, reconociendo diferencias básicas en las estrellas, pero sin establecer conexiones conceptuales entre estas características y su evolución Los resultados evidencian diferencias en la apropiación conceptual de los participantes, mostrando que los niveles más altos de comprensión se asocian con una mayor integración del contexto científico y sociocultural en el aprendizaje. La observación estelar nocturna no solo contribuye a la adquisición de conocimientos astronómicos, sino que también permite comprender su papel en la evolución del pensamiento humano, la navegación y el desarrollo tecnológico a lo largo de la historia.

**Análisis de los resultados del Taller 2: Espectro celeste visible**

El Taller 2 se centró en el estudio del espectro electromagnético visible y su relación con los cuerpos celestes. Su objetivo principal fue explorar la manifestación de la energía y las radiaciones electromagnéticas, enfocándose en la luz visible y su asociación con longitudes de onda específicas. Las actividades incluyeron la creación de un espectroscopio casero, la elaboración de infografías interactivas sobre la composición química de las estrellas, y experimentos para demostrar las propiedades de la luz para observar la reflexión, refracción y difracción de la luz. Las actividades promovieron el aprendizaje de conceptos astrofísicos mientras desarrollaban habilidades del PC como depuración, pensamiento algorítmico, automatización y abstracción.

*Análisis de las habilidades del pensamiento computacional (PC)*

Para el Taller 2, se evaluaron cuatro habilidades principales: abstracción (AB), pensamiento algorítmico (PA), automatización (AU) y depuración (DP), según las actividades diseñadas para el desarrollo del taller en mención (Figura 8).

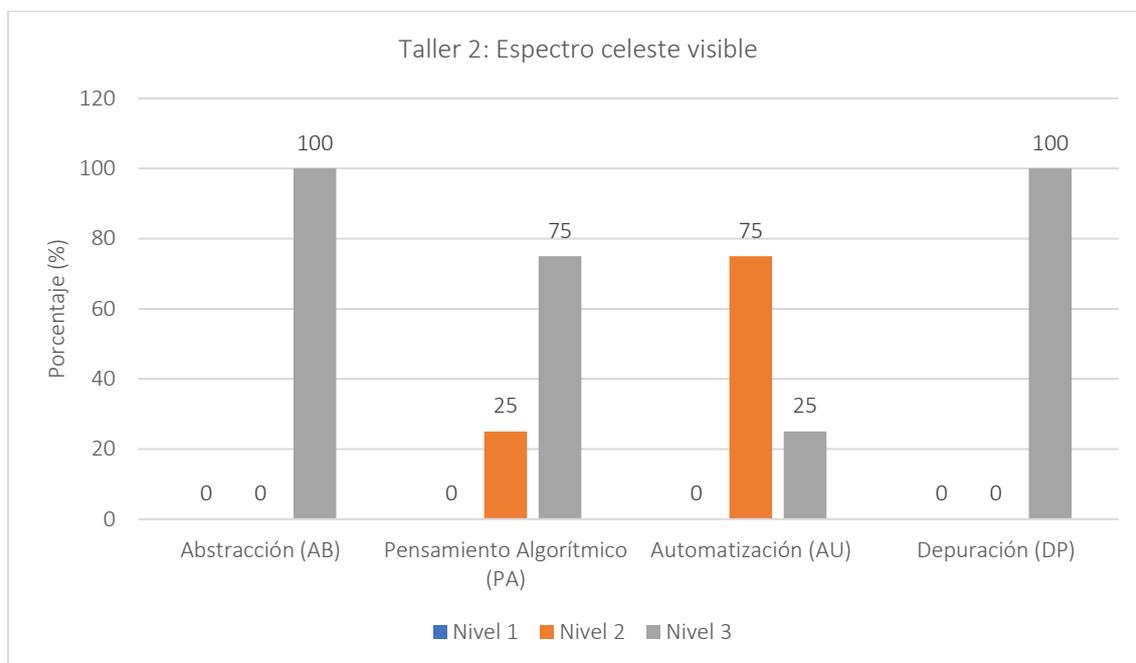

**Figura 8:** Niveles en las habilidades de pensamiento computacional (PC) del Taller 2.

En Pensamiento Algorítmico, el 25% de los grupos se ubicó en el nivel PA2, mientras que el 75% restante alcanzó el nivel PA3. Esto indica que la mayoría logró formular procedimientos detallados de manera adecuada, aunque algunos cometieron errores en la secuencia o aplicación. Esta habilidad permite estructurar soluciones mediante un conjunto de instrucciones organizadas paso a paso, facilitando la resolución de problemas en diferentes contextos (Pinzón y González, 2022). En el diseño del espectrofotómetro casero, los grupos que alcanzaron PA3 lograron un encadenamiento lógico de pasos sin errores significativos, mientras que los que quedaron en PA2 presentaron dificultades en la ejecución de la secuencia, lo que demuestra la importancia del pensamiento algorítmico en la planificación de procedimientos experimentales.

En Automatización, el 25% de los grupos alcanzó el nivel AU3, al crear infografías interactivas utilizando herramientas avanzadas para simplificar y organizar la información de manera dinámica. El 75% restante se ubicó en AU2, optando por infografías tradicionales, lo que indica un conocimiento básico de herramientas informáticas, pero con limitaciones en la automatización de procesos visuales. Las infografías interactivas facilitaron la identificación de parámetros relevantes y promovieron un acceso ágil a la información, permitiendo una mejor estructuración del contenido, tal como se había reportado en la literatura (Díaz-López, 2021). Este resultado es consistente con el enfoque de automatización en la enseñanza, donde la optimización de recursos digitales favorece la representación de conceptos complejos de forma intuitiva.

En Depuración, el 100% de los grupos alcanzó el nivel DP3, identificando y corrigiendo errores en los experimentos mediante análisis lógico y colaboración. Durante la ejecución de la actividad experimental, los participantes detectaron inconsistencias en sus mediciones y ajustaron variables como la intensidad de la luz o el ángulo de incidencia para mejorar la precisión de los resultados. Este proceso refleja la importancia de la depuración en entornos experimentales, donde la validación de resultados y la retroalimentación inmediata son esenciales para optimizar la metodología aplicada.

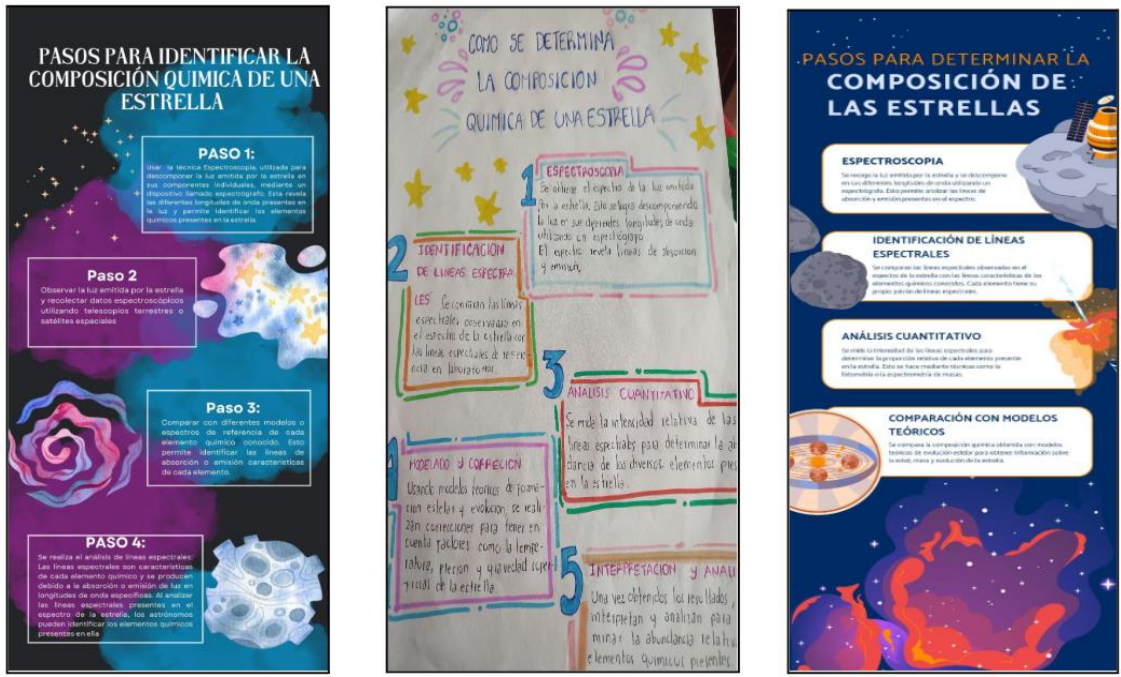

**Figura 9**: Infografías Grupo 3, Grupo 2 y Grupo 5

*Fortalecimiento conceptual*

Para el Taller 2, se analizó la apropiación conceptual de las estudiantes en temas relacionados con el espectro electromagnético visible, (Se incluyo en este la percepción de la luz visible), longitud de onda, frecuencia, y propiedades de la luz. Los resultados mostraron diversos niveles de comprensión alcanzados por los grupos (Figura 10).

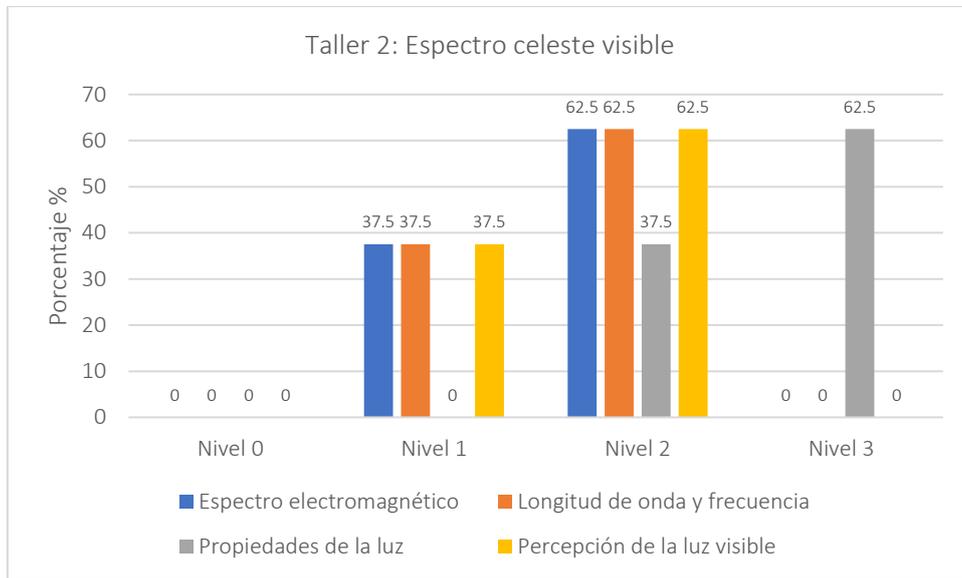

**Figura 10:** Niveles de apropiación conceptual y temática del Taller 2.

En relación con el concepto de espectro electromagnético visible, el 62.5% de los grupos lograron un nivel 2 de comprensión, asociando el espectro con las ondas visibles al ojo humano y estableciendo conexiones básicas con cuerpos celestes y la percepción de colores. No obstante, no profundizaron en las características fundamentales del espectro ni en su relación con los procesos físicos que generan la emisión de luz en distintas longitudes de onda. El 37.5% restante se ubicó en el nivel 1, mostrando respuestas superficiales y conocimientos básicos, lo que indica dificultades en la correlación entre la radiación electromagnética y su impacto en la percepción visual.

En cuanto a los conceptos de frecuencia y longitud de onda, el 62.5% de los grupos alcanzaron el nivel 2, estableciendo correlaciones entre estas variables y su relación con la intensidad de los colores y la energía, reconociendo la proporcionalidad inversa entre la frecuencia y la longitud de onda. Sin embargo, el 37.5% permaneció en el nivel 1, comprendiendo la diferencia entre frecuencia y longitud de onda, pero sin proporcionar ejemplos o explicaciones detalladas que evidenciaran una apropiación conceptual sólida.

Respecto a las propiedades de la luz (reflexión, refracción y difracción), el 37.5% de los grupos alcanzó el nivel 2, diseñando experimentos para representar estos fenómenos y ofreciendo explicaciones teóricas adecuadas. El 62.5% restante logró el nivel 3, demostrando una comprensión sólida y experimentando directamente estos fenómenos, lo cual fortaleció la conexión entre la experimentación y su contexto. La experimentación práctica no solo permitió validar los principios físicos subyacentes, sino que también favoreció la construcción de conocimiento significativo a partir de la observación y manipulación de variables. Como indica Trujillo (2018), la vivencia directa de los experimentos fomenta actitudes científicas y contribuye al aprendizaje significativo, permitiendo que los estudiantes enriquezcan su conocimiento al aproximarse a fenómenos cotidianos y estableciendo conexiones valiosas que favorecen una comprensión más profunda y duradera.

**Análisis de los resultados del Taller 3: Espectro electromagnético no visible**

En el Taller 3, enfocado en el espectro electromagnético no visible, particularmente en la región del infrarrojo, se llevaron a cabo actividades que incluyeron la exploración de conocimientos previos, experimentos prácticos y la construcción de circuitos para representar constelaciones. Estas actividades buscaban consolidar los conocimientos teóricos y establecer vínculos con aplicaciones tecnológicas cotidianas, promoviendo una integración práctica y significativa de los conceptos estudiados.

*Análisis de las habilidades del pensamiento computacional (PC)*

El análisis se centró en evaluar las habilidades del PC de abstracción (AB), pensamiento algorítmico (PA) y descomposición (D), a través de las actividades experimentales y de diseño de circuitos realizadas en el Taller 3 (Figura 11).

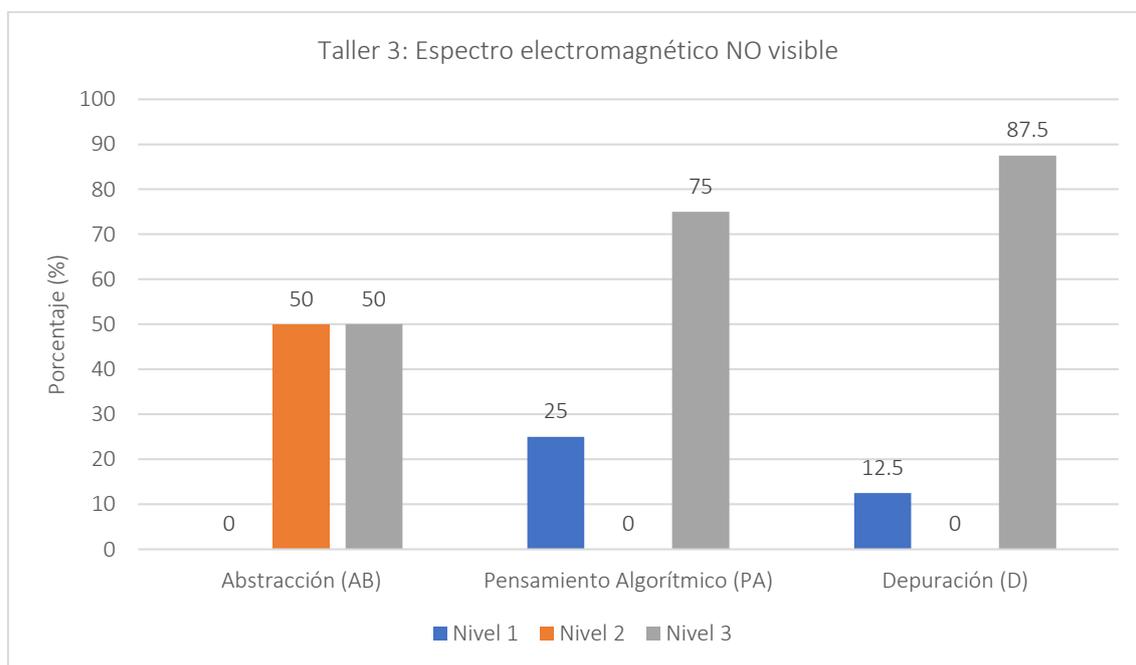

**Figura 11:** Niveles en las habilidades de pensamiento computacional (PC) del Taller 3.

En la habilidad de abstracción, el 50% de los grupos se encontró en el nivel intermedio (AB2), mostrando concepciones acertadas, pero sin argumentos detallados sobre el fenómeno de la luz infrarroja. El 50% de los grupos alcanzó el nivel más alto (AB3), al explicar de manera lógica y coherente por qué la cámara del celular podría captar la luz infrarroja mientras que el ojo humano no.

Respecto al pensamiento algorítmico, el 75% de los grupos se situó en el nivel más alto (PA3), presentando una estructura ordenada y detallada para la construcción de circuitos que representaban constelaciones. Ninguno de los grupos se ubicó en el nivel intermedio (PA2). El 25% de los grupos se encontró en el nivel más bajo (PA1), con dificultades para diseñar una metodología detallada en la construcción del circuito.

En cuanto a la descomposición, el 87.5% de los grupos logró alcanzar el nivel más alto (D3), al dividir el problema de la creación del circuito en tareas más sencillas, desde la elección de la constelación hasta la correcta conexión de los componentes (Figura 12). Ninguno de los grupos se ubicó en el nivel intermedio (D2). El 12.5% de los grupos se encontró en el nivel más bajo (D1), mostrando dificultades para especificar los componentes necesarios y la conexión de estos en el diseño y construcción del circuito.

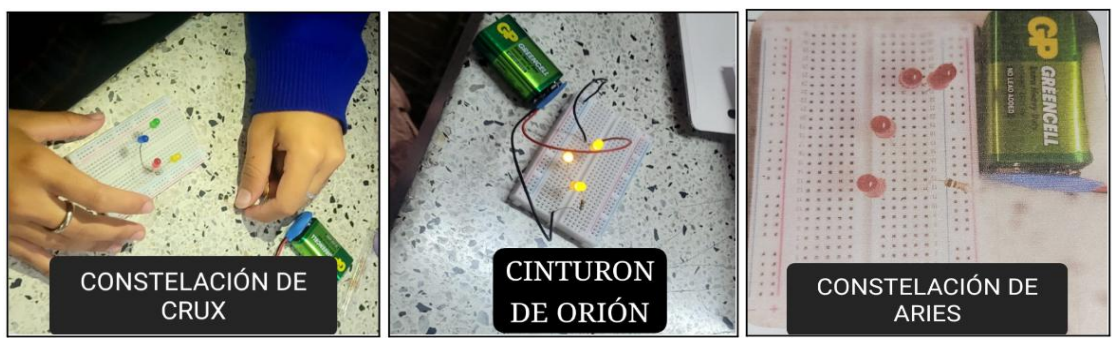

**Figura 12:** Constelaciones representadas mediante circuitos eléctricos.

El análisis reveló diferentes niveles de habilidades de PC en los grupos. La mayoría de los grupos alcanzó niveles altos en las tres habilidades evaluadas, especialmente en descomposición y pensamiento algorítmico. Sin embargo, en la habilidad de abstracción, hubo una división equitativa entre los niveles intermedio y alto. El análisis se enfocó en cómo estas habilidades se manifestaron en el diseño y ejecución de experimentos, confirmando que "el experimento es indispensable en la práctica, puesto que, permite una comprensión profunda en este tipo de fenómeno" (Brito, 2009).

*Fortalecimiento conceptual*

Para el taller 3, se analizó la apropiación conceptual de las estudiantes en temas relacionados con la región del espectro no visible, la región del infrarrojo y los circuitos eléctricos. Los resultados mostraron diversos niveles de comprensión alcanzados por los grupos (Figura 13).

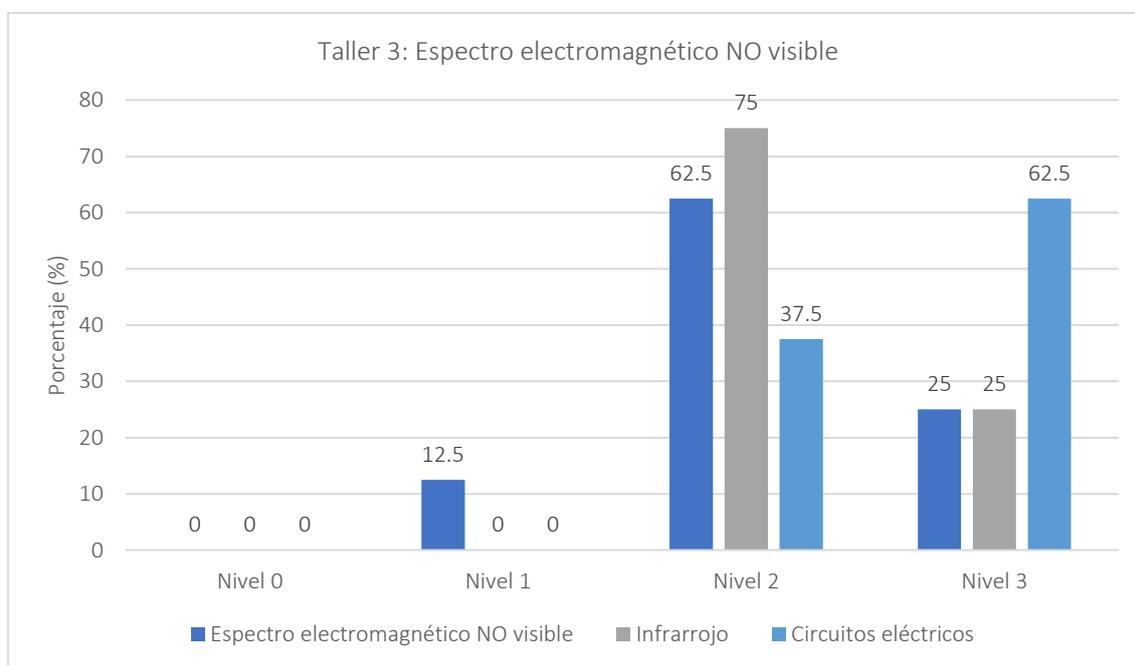

**Figura 13:** Niveles de apropiación conceptual y temática del Taller 3.

En cuanto al concepto de la región del espectro no visible, el 62.5% de los grupos se situó en el Nivel 2, analizando el espectro desde la longitud de onda y frecuencia, pero sin lograr relacionar completamente estos parámetros. Mencionaron algunas aplicaciones cotidianas, pero sin profundizar en ellas. El 25% alcanzó el Nivel 3, estableciendo una clara relación entre longitud de onda y frecuencia, y mencionando aplicaciones cotidianas del espectro no visible. El 12.5% restante se ubicó en el Nivel 1, reconociendo sólo las diferentes regiones del espectro electromagnético sin argumentar sus características.

Para el concepto de la región del infrarrojo, el 75% de los grupos alcanzaron el Nivel 2, analizando la región como inferior al espectro visible, pero sin establecer una clara relación entre longitud de onda y frecuencia. Identificaron algunas aplicaciones del infrarrojo, aunque de manera limitada. El 25% restante se situó en el Nivel 3, estableciendo la relación entre longitud de onda, frecuencia y aplicaciones tecnológicas del infrarrojo, incluyendo la detección de cuerpos celestes y temperatura de estrellas y planetas.

De acuerdo con los circuitos eléctricos, el 62.5% de los grupos se ubicó en el Nivel 3, demostrando una comprensión sólida al relacionar la Ley de Ohm en la construcción de circuitos, identificar la necesidad de resistencias para los LEDs y diseñar circuitos mixtos. El 37.5% restante alcanzó el Nivel 2, diseñando y ejecutando circuitos eléctricos sencillos, pero presentando falencias en la conexión y sin aplicar la Ley de Ohm.

Se observó una mejora en la comprensión y aplicación práctica de los conceptos, especialmente en circuitos eléctricos. Sin embargo, persisten algunas dificultades en la comprensión profunda de las características del espectro no visible y la región del infrarrojo. Esto subraya la importancia de continuar reforzando la enseñanza de estos conceptos, especialmente en su relación con fenómenos astronómicos y aplicaciones tecnológicas, para facilitar una comprensión más integral y su conexión con experiencias cotidianas y observaciones astrofísicas.

**Análisis de los resultados del Taller 4: Radioastronomía**

El Taller 4 se enfocó en la radioastronomía y su importancia en el estudio de objetos cósmicos, centrándose específicamente en la región de ondas de radio del espectro electromagnético. Las actividades incluyeron preguntas diagnósticas sobre radiotelescopios, un ejercicio de imaginación para diseñar un radiotelescopio, y la construcción práctica de un detector de ondas de radio utilizando materiales caseros, que permitieron a las estudiantes aplicar habilidades de pensamiento computacional como la abstracción (AB), el pensamiento algorítmico (PA) y generalización (G).

*Análisis de las habilidades del pensamiento computacional (PC)*

En la habilidad de abstracción, el 100% de los grupos alcanzó el nivel más alto (AB3), demostrando una excelente capacidad para identificar aspectos relevantes del caso Event Horizon Telescope y comprender la funcionalidad de un radiotelescopio.

Respecto al pensamiento algorítmico, el 62.5% de los grupos se situó en el nivel más alto (PA3), presentando una secuencia lógica y coherente en la construcción del radio casero y la resolución de problemas (Figura 15). El 37.5% de los grupos se ubicó en el nivel intermedio (PA2), mostrando algunas dificultades menores en el desarrollo secuencial de las actividades.

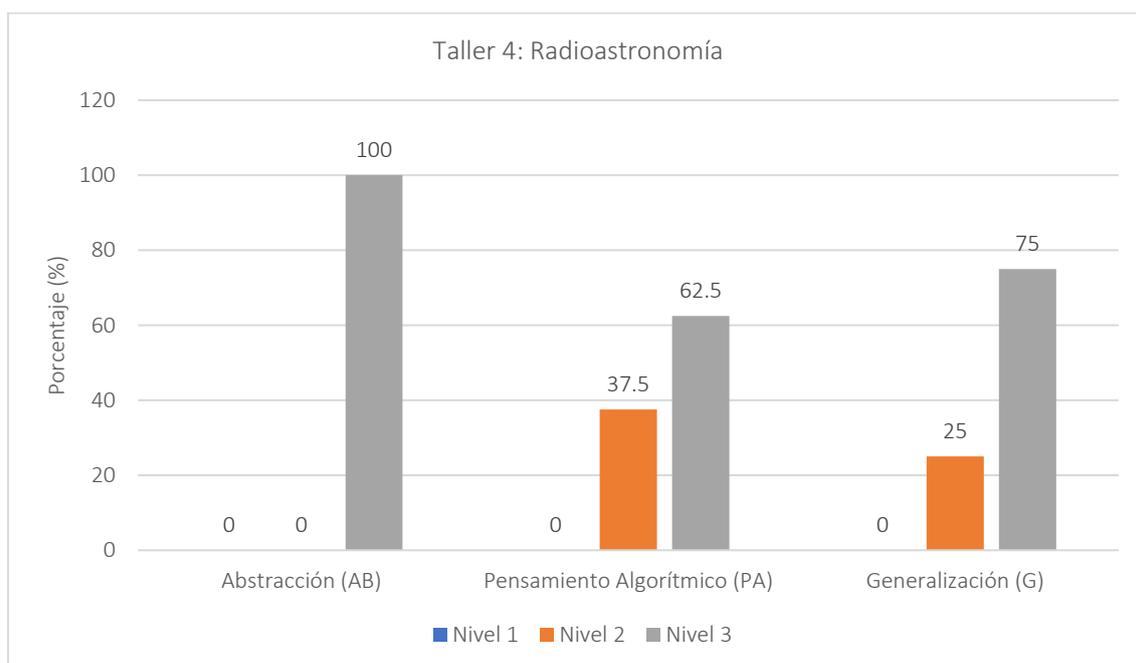

**Figura 14**: Niveles en las habilidades de pensamiento computacional (PC) del Taller 4.

En cuanto a la generalización, el 75% de los grupos logró alcanzar el nivel más alto (G3), demostrando una excelente capacidad para aplicar conceptos aprendidos en talleres anteriores y generar soluciones. El 25% restante se ubicó en el nivel intermedio (G2), mostrando una buena capacidad de generalización, pero con algunas limitaciones. Según Camargo y Munar (2021) esta habilidad genera soluciones sencillas de acuerdo con la identificación de concesiones y las relaciones existentes entre los problemas demostrando un entendimiento o solución profunda.

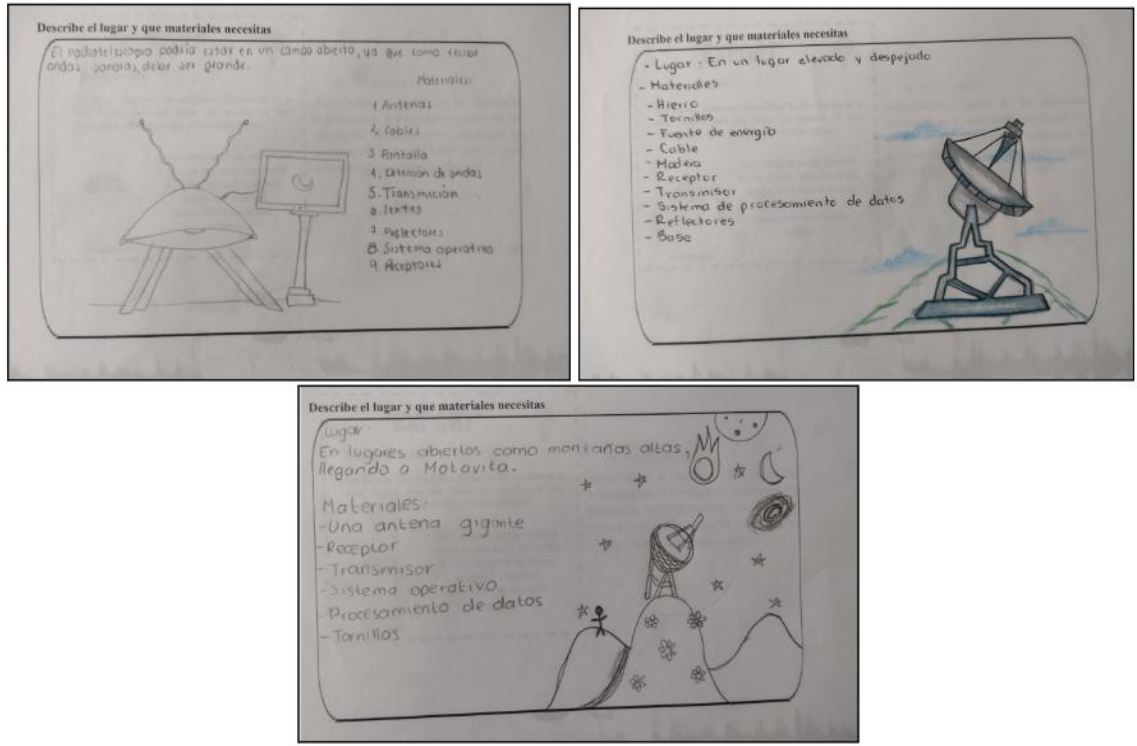

**Figura 15:** Diseño de radios

El análisis reveló un alto nivel de desarrollo de habilidades de PC en la mayoría de los grupos. Particularmente en abstracción, todos los grupos alcanzaron el nivel más alto, mientras que, en pensamiento algorítmico y generalización, la mayoría de los grupos se situaron en los niveles más altos. Esto permitió evaluar el progreso significativo en el desarrollo de estas habilidades a lo largo de la secuencia didáctica.

*Fortalecimiento conceptual*

Para el Taller 4, se analizó la apropiación conceptual de las estudiantes en temas relacionados con las ondas de radio y el electromagnetismo. Los resultados mostraron diversos niveles de comprensión alcanzados por los grupos (Figura 16).

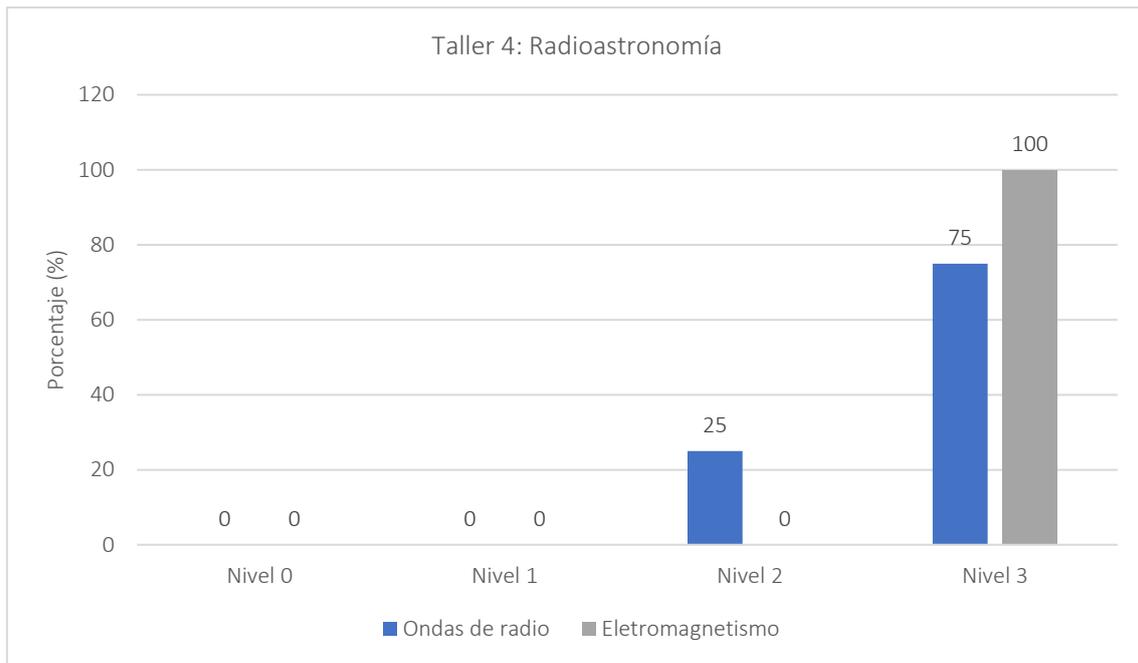

**Figura 16:** Niveles de apropiación conceptual y temática del Taller 4.

Para el concepto de ondas de radio, aproximadamente el 75% de los grupos alcanzó el Nivel 3, demostrando una comprensión profunda al explicar cómo las ondas de radio del espacio pueden ser captadas por radios comunes. El 25% restante se situó en el Nivel 2, ofreciendo definiciones no muy completas, sin profundizar en la relación entre las ondas de radio y el entendimiento del cosmos.

Respecto al electromagnetismo, el 100% de los grupos alcanzó el Nivel 3, reflejando una comprensión profunda de los fenómenos magnéticos y su relación con la radioastronomía. En general, se mejoró la comprensión y aplicación práctica de los conceptos de ondas de radio y electromagnetismo. Las respuestas de las estudiantes fueron más explícitas y ejemplificadas, ya que presentan un mayor vocabulario en comparación con respuestas de los primeros talleres.

Este taller demostró la eficacia de vincular conceptos de física con temas de astronomía. Como señala Pedraza (2020), "la enseñanza de las ondas de radio es esencial para la comprensión del electromagnetismo ya que se define dentro del aprendizaje de las ondas electromagnéticas" (p. 49). La actividad práctica de escuchar a Júpiter despertó la curiosidad de las estudiantes, fomentando la necesidad de indagar, preguntar y dar solución a problemas que surgen de cada fenómeno, motivando así la curiosidad e iniciativa por adentrarse al mundo de la física, mediante la astrofísica.

### Reflexiones generales

La estrategia didáctica implementada en esta investigación no solo demostró ser efectiva en el fortalecimiento de las habilidades de pensamiento computacional y la apropiación conceptual de la física y la astrofísica estelar, sino que también plantea oportunidades significativas para su aplicación en otros contextos educativos. La combinación de modelado computacional, experimentación y resolución de problemas permitió a los estudiantes establecer conexiones entre la teoría y la práctica, facilitando un aprendizaje más estructurado y significativo.

La adaptabilidad de esta estrategia sugiere su potencial para ser replicada en otras áreas del conocimiento dentro de la enseñanza de las ciencias naturales y la educación STEM (ciencia, tecnología, ingeniería y matemática) en general. Su aplicación en el estudio de otros fenómenos físicos, como la mecánica cuántica, la termodinámica o la óptica, podría contribuir a mejorar la comprensión de conceptos abstractos mediante herramientas computacionales y metodologías activas de aprendizaje. Asimismo, la incorporación del pensamiento computacional en niveles educativos más tempranos podría fortalecer el desarrollo de habilidades analíticas desde edades más tempranas, facilitando la transición a estudios más avanzados en física y astronomía.

Además, la contextualización de la estrategia en diferentes entornos socioeducativos, como instituciones rurales o comunidades con acceso limitado a laboratorios especializados, podría fomentar la democratización del conocimiento científico. El uso de herramientas accesibles, como simulaciones computacionales, dispositivos móviles y materiales de bajo costo, refuerza la viabilidad de este enfoque en diversos escenarios educativos, promoviendo una enseñanza de la ciencia más inclusiva y equitativa.

Futuras investigaciones podrían centrarse en la evaluación longitudinal del impacto de estas metodologías en la formación académica de los estudiantes y en su influencia en la toma de decisiones vocacionales en áreas científicas y tecnológicas. Asimismo, la integración de estrategias didácticas basadas en el pensamiento computacional en planes curriculares oficiales podría consolidar su papel como una herramienta pedagógica clave para el desarrollo de habilidades en la educación actual.

**Limitaciones y perspectivas de investigación**

Si bien la estrategia didáctica demostró ser efectiva en el desarrollo de habilidades de pensamiento computacional (PC) y la apropiación de conceptos en física y astrofísica estelar, su aplicación en un grupo específico de estudiantes limita su generalización. Futuras investigaciones podrían ampliar la muestra e incluir diferentes contextos educativos para evaluar su alcance.

La metodología utilizada permitió una intervención reflexiva, pero la influencia de factores externos, como el interés previo de los estudiantes o su familiaridad con herramientas computacionales, no fue completamente controlada. Diseños metodológicos mixtos fortalecerían el análisis del impacto del PC en el aprendizaje.

Además, la estrategia priorizó habilidades clave del PC, pero no profundizó en la optimización de algoritmos ni en simulaciones avanzadas. La integración de software especializado podría potenciar el análisis de datos astronómicos en futuras implementaciones.

Por otra parte, aunque la evaluación mostró avances en la comprensión conceptual y en el uso del PC, no se realizó un seguimiento longitudinal. Investigaciones futuras podrían analizar la consolidación de estas habilidades a lo largo del tiempo y su impacto en la formación académica y profesional de los estudiantes.

Estos aspectos abren nuevas posibilidades para fortalecer la enseñanza de la física mediante el PC, consolidando metodologías innovadoras que favorezcan un aprendizaje estructurado y aplicable a distintos contextos educativos.

## CONCLUSIONES

El desarrollo de habilidades de pensamiento computacional (PC) en la enseñanza de la física, particularmente en astrofísica, espectroscopía, circuitos eléctricos y radioastronomía, ha demostrado ser una estrategia eficaz para fortalecer la comprensión conceptual y el razonamiento lógico de los estudiantes. La combinación de enfoques experimentales y modelado computacional permitió estructurar soluciones a problemas científicos, mejorar la organización del pensamiento y potenciar la relación entre teoría y práctica.

Los resultados evidenciaron que los estudiantes mejoraron en la formulación de hipótesis, el uso de vocabulario técnico y la capacidad de estructurar problemas mediante algoritmos y simulaciones. Sin embargo, persisten desafíos en la automatización y la abstracción, lo que indica la necesidad de reforzar estas habilidades para lograr una integración más efectiva del PC en la enseñanza de la física. Además, la experimentación práctica y la aplicación del PC facilitaron la transición de los estudiantes de un aprendizaje descriptivo a un enfoque más analítico y estructurado.

A pesar de los beneficios observados, la aplicación del PC en la enseñanza de la física sigue siendo poco explorada en la literatura científica, con una predominancia de estudios centrados en informática y programación. Esto resalta la necesidad de ampliar la investigación sobre su implementación en la educación secundaria, explorando su impacto en la resolución de problemas científicos y en la formación de pensamiento crítico en contextos experimentales.

Se recomienda que futuras investigaciones profundicen en la relación entre PC y enseñanza de la física, abordando la optimización de estrategias didácticas que permitan una integración más equitativa de sus habilidades en distintos niveles educativos. Ampliar este enfoque contribuirá al diseño de metodologías innovadoras que favorezcan la alfabetización científica y tecnológica, preparando a los estudiantes para enfrentar los desafíos del siglo XXI desde una perspectiva interdisciplinaria y computacionalmente estructurada.

# REFERENCIAS

Bausela Herreras, Esperanza (2004). La docencia a través de la investigación-acción. *Revista Ibero Americana de educación, 35*(1), 1–9. https://doi.org/10.35362/rie3512871

Botella Nicolás, Ana; y Ramos Ramos, Pablo (2019). Investigación-acción y aprendizaje basado en proyectos. Una revisión bibliográfica. *Perfiles Educativos, 41*(163), 127–141. https://www.scielo.org.mx/scielo.php?script=sci_arttext&pid=S0185-26982019000100127#B49

Brito Ubaque, Karol Yobany (2009). Experimento: una herramienta fundamental para la enseñanza de la física. *Góndola, enseñanza y aprendizaje de las ciencias, 4*(1), 35-40. https://revistas.udistrital.edu.co/index.php/GDLA/article/view/5248/6886

Cadena Iñiguez, Pedro; Rendón Medel, Roberto; Aguilar Ávila, Jorge; Salinas Cruz, Eileen; De la Cruz Morales, Francisca; y Sangerman, Dora (2017). Métodos cuantitativos, métodos cualitativos o su combinación en la investigación: un acercamiento en las ciencias sociales. *Revista Mexicana de Ciencias Agrícolas, 8*(7), 1603-1617. https://doi.org/10.29312/remexca.v8i7.515

Camargo Pérez, Alvaro J.; y Munar Ladino, John A. (2021). Habilidades del pensamiento computacional en docentes en formación de la universidad La Gran Colombia. *Revista Científica UISRAEL, 8*(2), 135–149. https://doi.org/10.35290/rcui.v8n2.2021.441

Cristóbal Aragón, Esther (2017). *Desarrollo de habilidades de pensamiento mediante la enseñanza por indagación de contenidos de astronomía en primero de primaria* [Tesis de Grado, Universidad de Burgos]. Repositorio Institucional de la Universidad de Burgos. http://hdl.handle.net/10259/4548

Díaz-López, Mónica María (2021). Aprendizaje significativo de bioseguridad a través de infografías interactivas. *Educación Médica Superior, 35*(2). http://scielo.sld.cu/pdf/ems/v35n2/1561-2902-ems-35-02-e2736.pdf

González Pardo, Lorena M.; y Valderrama, Daniel A. (2021). Enseñanza de la física en pandemia; una experiencia desde el enfoque CTS. *Revista Tecné, Episteme y Didaxis*: TED, 274-280. https://core.ac.uk/reader/483511230

Gurdián-Fernández, Alicia (2010). *El paradigma cualitativo en la investigación socio educativa*. Coordinación Educativa y Cultural Centroamericana (CECC) y Agencia Española de Cooperación Internacional (AECI). PrintCenter, San José, Costa Rica. ISBN: 978-9968-818-32-2.

http://ice.ua.es/ar/documentos/recursos/materiales/el-paradigma-cualitativo-en-la-investigacion-socio-educativa.pdf

Hernández Arteaga, Isabel (2012). Investigación cualitativa: una metodología en marcha sobre el hecho social. *Rastros rostros, 14*(27), 57-68. https://revistas.ucc.edu.co/index.php/ra/article/view/444

Hernández, L., Torrero, P., & Hernández, V. (2014). Mapas mentales, mapas conceptuales, diagramas de flujo, esquemas. Red Durango de Investigadores Educativos, A.C.

Justiniano, Artur; y Botelho, Rafael (2016). Construção de uma carta celeste: Um recurso didático para o ensino de Astronomia nas aulas de Física. *Revista Brasileira de Ensino de Física, 38*. https://doi.org/10.1590/1806-9126-RBEF-2016-0131

Martínez Raba, Duván; y Moreno Católico, Angie (2024). *Pensamiento Computacional: Un Abordaje Didáctico Para El Aprendizaje De La Astrofísica ESTELAR* [Tesis de Grado, Universidad Pedagógica y Tecnológica de Colombia]. ResearchGate. http://dx.doi.org/10.13140/RG.2.2.21082.15043

National Research Council (2011). *A Framework for K-12 Science Education: Practices, Crosscutting Concepts, and Core Ideas.* Committee on a Conceptual Framework for New K-12 Science Education Standards. Board on Science Education, Division of Behavioral and Social Sciences and Education. Washington, DC: The National Academies Press. https://doi.org/10.17226/13165

Parra Zeltzer, Víctor; Vanegas Ortega, Carlos; y Bustamante González, Denisse (2021). La clase de física es una extensión de la clase de matemática: percepciones de estudiantes de enseñanza media sobre la enseñanza de la física. *Estudios Pedagógicos (Valdivia), 47*(3), 291–302. https://doi.org/10.4067/S0718-07052021000300291

Pedraza Ahumada, Leidy L. (2020). *Una mirada diferente al espacio: propuesta de enseñanza hacia las ondas de radio por medio de la observación astronómica partiendo de las ondas electromagnéticas* [Tesis de grado, Universidad Pedagógica y Tecnológica de Colombia] Archivo digital. http://hdl.handle.net/20.500.12209/12461

Percy, John (2012). *Evolución de las estrellas*. Publicaciones de NASE. *https://amyd.quimica.unam.mx/pluginfile.php/6406/mod_resource/content/1/evolucion%20de%20estrellas.pdf*

Pinzón Pérez, Diego F.; y González Palacio, Enoc V. (2022). Incidencia de las habilidades de pensamiento algorítmico en las habilidades de resolución de problemas: una propuesta didáctica en el contexto de la educación básica secundaria. *Estudios pedagógicos (Valdivia), 48*(2), 415-433. http://dx.doi.org/10.4067/S0718-07052022000200415

Rodríguez, Jenny Marcela (2018). Polarización de la luz: conceptos básicos y aplicaciones en astrofísica. *Revista Brasileira de Ensino de Física, 40*(4), e4310. https://doi.org/10.1590/1806-9126-RBEF-2018-0024

Rojas Miranda, Audis; y Vidal Herrera, Juan (2022). *Estrategia didáctica basada en la metodología STEAM y el aprendizaje basado en problemas -ABP- para la enseñanza de las ciencias en la Institución educativa San José del Pantano- Puerto Escondido-Córdoba* [Tesis de Grado, Universidad de Córdova]. https://repositorio.unicordoba.edu.co/bitstream/ucordoba/6472/1/rojasmirandaaudis-vidalherrerajuan.pdf

Surroca Carrascosa, Alfredo (2019). Azarquiel, el astrolabio y la azafea. Su aportación a la ciencia astronómica ya la navegación del Renacimiento. *Boletín de la Real Sociedad Geográfica, (CLIV)*, 115-138. https://boletinrsg.com/index.php/boletinrsg/article/view/81>

Triana González, Yaneth R. (2024). Formación de docentes en astronomía para mejorar el pensamiento científico y crítico en los estudiantes. *Tecné, Episteme y Didaxis: TED, 55*, 940–944. https://revistas.upn.edu.co/index.php/TED/article/view/21028

Trujillo Fernández, Gara (2018). Una aventura hacia la luz y sus propiedades. [Tesis de pregrado, Universidad de La Laguna]. Repositorio Institucional. http://riull.ull.es/xmlui/handle/915/8992

Unzueta Morales, Sandra (2011). Algunos aportes de la psicología y el paradigma socio crítico a una educación comunitaria crítica y reflexiva. *Revista Integra Educativa 4*(2), 105-144. http://www.scielo.org.bo/scielo.php?script=sci_arttext&pid=S1997-40432011000200006&lng=es&tlng=es

Valderrama, Daniel A. (2025). *Cualificación para la enseñanza de la astronomía en el contexto de la formación inicial docente* [Tesis de Maestría, Universidad Pontificia Bolivariana]. Respositorio DSpace. https://repository.upb.edu.co/handle/20.500.11912/12312

Valderrama, Daniel A.; y González Pardo, Lorena M. (2024). Transmutando la Inercia Pedagógica: Aprendizaje Activo en la Enseñanza de la física Transmuting Pedagogical Inertia: Active Learning in Physics Teaching. *Opuntia Brava, 16*(4). https://www.researchgate.net/publication/385379958

Vallejo Villegas, Agustín (2022). *Creación e Implementación Didáctica de Simulaciones Interactivas para la Enseñanza de Fenómenos Astrofísicos*. [Tesis de Grado, Universidad de Antioquia]. https://agusvallejo.art/assets/A.%20Vallejo-Villegas%20(2022).pdf

Vera Vélez, Lamberto (2008). *La Rúbrica y la Lista de Cotejo. Universidad Interamericana de Puerto Rico, Recinto de Ponce*, Departamento de Educación y Ciencias Sociales. www.quadernsdigitals.net/index.php?accionMenu...tipo

Verdejo Paris, Pilar (2008). *Modelo para la Educación y Evaluación por Competencias (MECO). Propuestas y acciones universitarias para la transformación de la educación superior en América Latina. Informe final del Proyecto 6x4 UEALC, 155-195.* http://www.innovacesal.org/redic_v2016/proyecto_6x4/p01/6x4_p01c.pdf

Wing, Jeannette (2006). Computational Thinking. *Communications of the ACM. 49*, 33-35. https://www.cs.cmu.edu/~15110-s13/Wing06-ct.pdf